\documentclass[useAMS,usenatbib]{mn2e}
\usepackage{graphicx}
\usepackage{lscape}

\title[Inverted spectrum extragalactic radio sources]
  {Extragalactic radio sources with sharply inverted spectrum
at metre wavelengths}
\author[Gopal-Krishna et al.]
  {Gopal-Krishna$^{1}$\thanks{E-mail:gopaltani@gmail.com},
   S. K. Sirothia$^{2,3,4}$,
   Mukul Mhaskey$^{5}$,
   Pritesh Ranadive$^{5}$,
\newauthor
   Paul J. Wiita$^{6}$,
   A. Goyal$^{7}$,
   N. G. Kantharia$^{2}$ and
\newauthor
   C. H. Ishwara-Chandra$^{2}$
\\
$^1$ Inter University Centre for Astronomy \& Astrophysics, Pune University 
Campus, Pune 411 0007, India\\
$^2$ National Centre for Radio Astrophysics/TIFR, Pune University Campus, 
Pune 411 007, India\\ 
$^3$ Square Kilometre Array South Africa, 3rd Floor, The Park, Park Road,
Pinelands, 7405, South Africa\\
$^4$ Department of Physics and Electronics, Rhodes University, PO Box 94,
Grahamstown, 6140, South Africa\\
$^5$ Department of Physics, University of Pune, Pune 411007, India\\
$^6$ Department of Physics, The College of New Jersey, PO Box 7718, Ewing, 
NJ 08628, USA\\
$^7$ Astronomical Observatory, Jagiellonian University, ul. Orla 171, 30-244, Krakow, Poland}
\date{Accepted 2014 July 6.  Received 2014 July 1; in original form 2013 October 31}

\pagerange{\pageref{firstpage}--\pageref{lastpage}} \pubyear{2013}

\def\LaTeX{L\kern-.36em\raise.3ex\hbox{a}\kern-.15em
    T\kern-.1667em\lower.7ex\hbox{E}\kern-.125emX}

\begin{document}
\label{firstpage}

\maketitle

\begin{abstract}

We present the first results of a systematic search for the rare extragalactic 
radio sources showing an inverted  (integrated) spectrum, with spectral index 
$\alpha \ge +2.0$, a previously unexplored spectral domain. The search is expected
to yield strong candidates for $\alpha \ge +2.5$, for which the standard synchrotron self-absorption
(characterized by a single power-law energy distribution of relativistic electron population)
would not be a plausible explanation, even in an ideal case of a 
perfectly homogeneous source of incoherent synchrotron radiation. 
Such sharply inverted spectra, if found, would require alternative explanations, 
e.g., free-free absorption, or non-standard energy distribution of relativistic
electrons which differs from a power-law (e.g., Maxwellian).

The search was carried out by comparing two sensitive low-frequency radio 
surveys made with sub-arcminute resolution, namely, the WISH survey at 352 MHz and 
TGSS/DR5 at 150 MHz. The overlap region between these two surveys contains 7056 WISH 
sources classified as `single' and brighter than 100 mJy at 352 MHz. We focus here 
on the seven of these sources for which we find $\alpha > +2.0$. Two of these are undetected at 
150 MHz and are particularly good candidates for $\alpha > +2.5$. Five of the 
seven sources exhibit a `Gigahertz-Peaked-Spectrum' (GPS).

\end{abstract}


\section{Introduction}


\vskip0.5in

Synchrotron radiation that dominates the radio-frequency output 
of radio galaxies arises from regions covering a vast range 
in spatial scale.  A warm gaseous environment around some or 
all of these components can, in principle, produce visible 
signatures (via free-free absorption: FFA) in the form of an 
inverted radio continuum spectrum. However, inverted 
radio spectra, when observed, are most commonly interpreted in 
terms of synchrotron self-absorption (SSA) occurring within the 
parsec-scale, or even more compact radio components revealed 
by Very Long Baseline Interferometry (VLBI) (e.g., Urry \& 
Padovani 1995; O'Dea \& Baum 1997). This is mainly because the standard
SSA, which is characterized by a single power-law energy distribution
of relativistic electrons in the source, can by itself produce an inverted spectrum with $\alpha$ as
large as $\alpha_c$ = +2.5 ($S_{\nu} \propto \nu^{\alpha}$). Although attainable 
only for an implausibly ideal, perfectly homogeneous radio source, this 
spectral index limit is practically independent of the slope of 
power-law energy distribution of the radiating relativistic electrons  
(e.g., Slish 1963, Rybicki \& Lightman 1979). 
As argued by Rees (1967), SSA can 
even cause $\alpha$ somewhat larger than $\alpha_c$ over just about 
one decade in frequency, when a significant non-power-law (e.g., 
mono-energetic, or Maxwellian) component is also present in the energy distribution of the
relativistic electron population. A similar conclusion was reached 
by de Kool \& Begelman (1989) by considering energy distributions in 
which the number of low-energy electrons is over-represented 
relative to a power-law extrapolation of the distribution at higher
energies. Thus, finding examples of synchrotron radio sources 
having $\alpha > \alpha_c$ = +2.5 would help in 
tracing rare extragalactic radio sources dominated by either non-standard 
synchrotron emission process or FFA effects.  

Until now, $\alpha > +2.5$ has only been demonstrated for 
parsec-scale inner segments of the VLBI jets in a few radio galaxies 
and quasars. Examples include the well known sources 
NGC 1275/Perseus A (Walker et al.\ 2000; Levinson et al.\ 1995), 
NGC 4261 (Jones et al.\ 2001) and 3C 345 (Matveenko, Pauliny-Toth 
\& Sherwood 1990). In these cases, the ultra-steep inverted radio 
spectra have generally been attributed to FFA. Likewise, in the nearby 
elliptical NGC 1052, the innermost portion of the sub-parsec scale 
counter-jet shows a spectral cut-off with $\alpha$ becoming larger 
than $+3$, which too has been interpreted in terms of FFA arising 
from an annular ring of ionized gas, possibly a geometrically thick, 
patchy disc (e.g., Vermeulen et al.\ 2003; also Kadler et al.\ 2004).  

As far as the {\it integrated} radio spectra are concerned, an 
outburst with an extremely inverted spectrum with $\alpha = 
+1.9\pm0.1$ between 4.8 and 10.5 GHz has been reported for the 
blazar-like spiral galaxy III Zw 2 (Falcke et al.\ 1999) and it was 
attributed to SSA. Note that the integrated radio spectrum in this 
case was dominated by the flare and was thus a transient
feature. More recently, Murphy et al.\ (2010) have noted a subset 
of extragalactic radio sources they termed `Ultra-inverted Spectrum 
Radio Sources' (UIS). They showed that nearly $0.4$ per cent of the 
sources detected in the Australia Telescope Compact Array 20 GHz 
survey (AT20G) lack a counterpart in the lower frequency catalogues, 
like the NRAO Very Large Array Sky Survey (NVSS) at 1.4 GHz (Condon et al.\ 1998) 
and the Sydney University Molonglo Sky Survey (SUMSS)/ Molonglo Galactic 
Plane Survey (MGPS-2) at 843 MHz (Mauch et al.\ 2003; Murphy et al.\ 2007). 
The implied spectral indices between 5 and 20 GHz typically are 
$\alpha \simeq +0.7$, although values approaching +2.0 are also observed 
in a few cases; these are still consistent with the standard SSA 
characterized by a single power-law energy distribution of relativistic 
electrons . A similar situation exists in the case of `High-Frequency-Peakers' (HFPs) 
studied by Dallacasa et al.\ (2000).

One numerically significant class of sources for which a low-frequency spectral turnover 
is the defining characteristic is the `Gigahertz-Peaked-Spectrum' 
(GPS) sources (e.g., Gopal-Krishna, Patnaik \& Steppe 1983; Spoelstra,
Patnaik \& Gopal-Krishna 1985; Gopal-Krishna \& Spoelstra 1993). 
The earliest example of such sources can be traced back to Bolton, Gardner \& Mackey (1963) and
Kellermann (1966) (see, also, Phillips \& Mutel 1982). These sources 
exhibit a convex radio spectrum peaking in the region between 1 
and 5 GHz, are mostly $< 1$ kiloparsec in size and make up nearly 
10 per cent of the extragalactic population found in high-frequency radio 
surveys (reviewed, e.g., in O'Dea 1998; R.\ Fanti 2009; C.\ Fanti 2009; 
Tzioumis et al.\ 2003). Although their spectra are rarely well determined at 
frequencies below the turnover, the standard SSA again continues to be the 
popular explanation for the spectral turnover (e.g., O'Dea \& Baum 1997; 
Snellen et al. 1999). Nonetheless, dominance of FFA 
has also been considered for almost half a century (e.g., Kellermann 1966; van Breugel 1984; Bicknell et al.\ 1997; 
Kuncic et al.\ 1998; Begelman 1999; Kameno 
et al.\ 2000; Marr, Taylor \& Crawford 2001; Shaffer, Kellermann \& Cornwell 1999; 
Tingay \& de Kool 2003; Stawarz et al.\ 2008; Ostorero et al.\ 2010).   
In case FFA occurs in a uniform ambient medium, the spectrum of 
the synchrotron radio source at lower frequencies would exhibit an exponential cut-off,
eventually becoming steeper than $\alpha_c = +2.5$. But, in practice, 
the slope of the inverted spectrum would significantly depend on factors 
like $\alpha$ at frequencies where the synchrotron source is transparent 
and the distribution of free-free optical depth within the (clumpy) ambient 
thermal plasma. Consequently, FFA interpretation cannot be excluded 
even if the inverted spectrum has a slope well under $\alpha_c = +2.5$ (e.g., Bicknell et al.1997; 
Kellermann 1966). 

\section{Search for the extreme spectral turnovers}  

It is clearly of interest to find out if extragalactic radio 
sources with  $\alpha \ge +2.5$ do exist. As mentioned above, observations 
reported so far have shown that even radio spectra approaching 
$\alpha = +2.0$ are exceedingly rare. The present work is an attempt to 
explore the next steeper spectral regime,  i.e., the one characterized by
$\alpha > +2.0$. We shall term the extragalactic radio sources falling 
within this extreme spectral domain as 
{\it `Extremely Inverted Spectrum Extragalactic Radio Sources: EISERS'}. 
Here we present the first results of a sensitive search for this rare population, 
made by comparing the 352 MHz `Westerbork In the Southern Hemisphere' 
(WISH) survey (de Breuck et al.\ 2002) and the ongoing TIFR.GMRT.SKY.Survey 
(TGSS) at 150 MHz\footnote{http://tgss.ncra.tifr.res.in/} 
using the Giant Metrewave Radio Telescope\footnote{Operated by the 
National Centre for Radio Astrophysics (NCRA) of the Tata Institute 
of Fundamental Research (TIFR)} (GMRT, Swarup et al. 1991). Both the 
WISH survey and the portion of TGSS reported so far are confined to 
negative declinations where they not only are the deepest large-area sky 
surveys available at metre-wavelengths, but also have a fairly high 
(sub-arcminute) angular resolution. Due to these key advantages, they 
are well suited for picking the most promising candidates for 
extragalactic sources having inverted radio spectra with a slope 
$\alpha > $ $\alpha_c = $+2.5 discussed above. Such steep radio spectra 
may even be confirmed for some of the sources for which present 
estimate of $\alpha$ (150-352 MHz) is slightly less extreme (+2.0 $< \alpha < $+2.5), 
when flux densities are measured with higher precision 
and, moreover, the effect of flux variability are minimised by making 
multi-frequency observations quasi-simultaneously. 

The WISH survey was made with the Westerbork Synthesis Radio Telescope 
(WSRT) with a FWHM of 54" x 54" cosec ($\delta$) and a typical 
rms noise of 3.5 mJy/beam at 352 MHz (for details, see de Breuck et al.\ 2002). 
Likewise, the TGSS is being carried out using the Giant Metrewave 
Radio Telescope (GMRT) with a FWHM of approximately 20 arcsec and 
a typical rms noise of 6 mJy/beam at 150 MHz. The 5th data release
of TGSS (DR5) covers a little more than a steradian of the sky  at negative 
declinations down to $-35^\circ$. It consists of 558 sky frames of size $\sim$ 
4.5$^\circ$ x 4.5$^\circ$ each.
The source positions in both TGSS and WISH catalogues are tied to the NVSS 
coordinate frame. Near the $90\%$ completeness limit of the TGSS 
survey ($\sim 40$ mJy at 150 MHz, for unresolved sources), the rms 
positional accuracy is $\sim 4$ arcsec. The flux density errors are 
expressed in terms of rms noise measured in the proximity of a given 
source (see below). Since a sky map of brightness temperature is not 
yet available at 150 MHz (this work is underway, using GMRT) we have 
re-calibrated, for the present purpose, the flux densities of individual 
sources given in the TGSS/DR5 catalogue. For this, we first determined 
spectral indices of all 52839 'single component sources' in the 
WISH catalog (designated type `S') by comparing
their flux densities at 352 MHz with those of their counterparts found 
(within 30 arcsec) in the NVSS catalogue at 1.4 GHz. This gave a $\it median$ 
value of $\alpha$ (352-1400 MHz) = -0.748 $\pm$ 0.001, which we then 
assumed to extend down to 150 MHz and thus be applicable to the sources 
detected in each TGSS frame of extent 4.5$^\circ$ x 4.5$^\circ$  
(see below). 
Satisfying this requirement of median $\alpha$ (150-1400 MHz) = -0.75 
for each TGSS frame yielded a `flux scaling factor' (FSF) for that frame at 150 MHz, 
by which the catalogued flux density of each source in that frame, as 
well as its quoted rms error should be multiplied. Note that the rms flux density
error given for a source in the TGSS catalogue actually refers to a
region within 4 arcmin of the source (but excluding the source itself), 
so it includes a source flux density dependent term (for details, see 
Sirothia et al.\ 2009). Thus, the rms flux density error for a source, as 
given in the TGSS catalogue, once multiplied with the FSF determined 
for the corresponding TGSS frame, amounts to the true rms error of the 
flux density of that source at 150 MHz (see Table 1). 

Considering only the type `S' (i.e. single component) WISH sources having 
integrated flux densities above 100 mJy at 352 MHz, their 352 MHz flux 
densities were combined with their TGSS/DR5 flux densities at 150 MHz 
(or, upper limits in case of non-detections) scaled using the above 
determined values of FSF for the respective TGSS frames. The resulting 
values of $\alpha$ (150-352 MHz) were used to shortlist steeply inverted 
spectrum sources with $\alpha$ (150-352 MHz) $>$ +1.75. The next
step was to assess the reliability of these $\alpha$ estimates, paying attention to 
the rather large north-south extent of the WISH survey beam 
(54" x 54" cosec $\delta$). To do this, we examined for each source its 
higher resolution images available in the TGSS and NVSS and whenever 
an image was found to be clearly resolved, or having neighbourhood 
radio emission that could have contaminated the (lower resolution) WISH 
measurement at 352 MHz, the source was discarded from further consideration. 
Further, among the retained WISH sources, if a TGSS counterpart was not detected, 
we have set for its 150 MHz flux density an upper limit equal to 2.5 times the rms 
noise for that source, as given in the TGSS/DR5 catalog and scaled using 
the FSF estimated for the corresponding TGSS frame, as described above.

\section{Results and concluding remarks} 

Table 1 lists the seven sources found here to have $\alpha$ (150-352 MHz) $>$ 
+2.0. Out of these, two sources are undetected at 150 MHz and therefore only 
lower limits to their $\alpha$ values are provided. As mentioned above, 
all TGSS flux densities at 150 MHz have 
been rescaled here using the FSF scaling factors
determined for individual TGSS frames, based on the assumption 
that for each frame the median value of source spectral indices, determined between 352 MHz and 1400 MHz, holds 
down to 150 MHz (Sect. 2). This is a conservative approach because it is known 
that, on average, metre/decimetre-wave spectra of extragalactic radio 
sources tend to flatten with increasing wavelength (e.g., Laing, Riley \& Longair 1983; 
Mangalam \& Gopal-Krishna 1995). Thus, due to our assumption of 
straight spectrum down to 150 MHz, the values of  $\alpha$(150-352 MHz) of individual sources, 
as estimated here should in fact be slightly on the lower side 
(i.e., indicating less sharply inverted spectra than are actually present). 

The seven sources reported here (Table 1) are the first examples of EISERS and by far 
the best available candidates to look for an inverted spectrum with
$\alpha > \alpha_c = +2.5$, requiring either FFA, or a non-standard SSA. 
For the two best candidates, which are undetected at 150 MHz, 
the TGSS and NVSS images are shown in Figure 1. Optical counterparts for all 
seven sources were searched in the NED database and the 
Digital-Sky-Survey (DSS) database and they were found for two of the 
sources (Table 1). Deeper optical imaging is needed to identify the 
remaining five EISERS.

All the seven EISERS appear unresolved in their NVSS/TGSS 
images. For some of them, the NED database provides flux densities at 
gigahertz frequencies, based on high-quality measurements made with 
sub-arcminute beams. We have included these data in Table 1 and shown
the resulting radio spectra in Figure 2. At least five of the seven spectra 
are found to be of GPS type (Sect. 1). However, a fully secure radio spectral 
characterization would need nearly simultaneous multi-frequency flux 
measurements. It may be noted that in the present case, the flux 
measurements at these low frequencies (TGSS) were made almost a decade after 
those made at 352 MHz (WISH) and hence some of the computed spectral 
indices may have been significantly affected by flux variability, although 
the probability of a substantial variation at these low frequencies is minute, specially for 
GPS sources (e.g., O'Dea 1998). For general population of extragalactic 
radio sources, flux variability at 150 MHz on year-like time scale has been 
investigated in a few programmes, as summarized by Bell et al. (2014). 
In a sample of 811 relatively faint sources (flux density $>$ 0.3 Jy at 151 MHz), 
observed by McGilchrist \& Riley (1990), no significant fractional variability
above $4\%$ was found; at higher flux levels only about $2\%$ 
of the sources are reported to vary by more than $10\%$ at 154 MHz (the
variability could either be intrinsic, or due to propagation effects, cf. 
Bell et al. 2014). Nonetheless, given that sources with $\alpha > +2.5$ 
must be exceedingly rare, quasi-simultaneous, multi-frequency flux  
measurements at metre/decametre wavelengths would be particularly 
desirable for these seven sources and additional candidates for $\alpha > +2.5$
that our ongoing search is expected to reveal.

\section*{Acknowledgments}
The Giant Metrewave Radio Telescope (GMRT) is a national facility operated 
by the National Centre for Radio Astrophysics (NCRA) of the Tata Institute 
of Fundamental Research (TIFR).  We thank the staff at NCRA and GMRT for 
their support. We also wish to thank an anonymous referee for the constructive 
criticism of the original version of this paper. This research has used the 
TIFR. GMRT. Sky. Survey (http://tgss.ncra.tifr.res.in) data products, 
NASA's Astrophysics Data System and NASA/IPAC Extragalactic Database (NED), 
which is operated by the Jet Propulsion Laboratory, California 
Institute of Technology, under contract with National Aeronautics and 
Space Administration.


\newpage
\begin{landscape}
\begin{table}
\small
\caption{Flux densities (mJy) and other radio properties of the seven EISERS}
\begin{tabular}{ccccccccc}\\
\hline

\multicolumn{1}{c}{Source position} & Optical ID & \multicolumn{1}{c}{150MHz}$^{\#}$ & \multicolumn{1}{c}{352MHz}$^{\dag}$ & \multicolumn{1}{c}{1.4GHz} & \multicolumn{1}{c}{4.85GHz} & \multicolumn{1}{c}{8GHz} & \multicolumn{1}{c}{20GHz} & \multicolumn{1}{c}{Spectral Index} \\
NVSS$^{\star}$ & redshift (NED)$^{\star}$&TGSS$^{\star}$ & WISH$^{\star}$ & NVSS$^{\star}$ & PMN$^{\star}$ & ATCA$^{\star}$ & ATCA$^{\star}$ & (150-352 MHz) \\

\hline
{\bf Strong Cases}{$^{\P}$} & & & & & & & & \\
\\
$ $02 42 10.64 & & $<$14.24 & 106${\pm}$4.5 & 96.7${\pm}$2.9 & & & & $>$2.35  \\
$-$16 49 32.9 & & & & & & & & \\
\\
$ $12 09 14.65 & G & $<$27.69 & 207${\pm}$8.4 & 353.7${\pm}$10.6 & 573${\pm}$32 & 1177${\pm}$61 & 707${\pm}$46 & $>$2.36  \\
$-$20 32 39.9 & z=0.404 & & & & & & & \\

\\
{\bf Probable Cases} & & & & & & & & \\
\\
$ $04 42 01.24 & & 16.33${\pm}$7.3 & 105${\pm}$4.4 & 50.9${\pm}$1.6 & 85${\pm}$11 & & & 2.18${\pm}$0.53  \\
$-$18 26 33.6 & & & & & & & & \\
\\
$ $10 03 06.11 & & 17.51${\pm}$4.1 & 143${\pm}$6.1 & 74.1${\pm}$2.8 & & & & 2.46${\pm}$0.32 \\
$-$25 14 04.3 & & & & & & & &  \\
\\
$ $10 31 52.36 & G & 30.06${\pm}$5.9 & 191${\pm}$7.9 & 371.8${\pm}$11.2 & 328${\pm}$20 & 291${\pm}$15 & 124${\pm}$8 & 2.17${\pm}$0.24 \\
$-$22 28 23.4 &  & & & & & & & \\
\\
$ $12 07 06.05 & & 67.08${\pm}$8.2 & 380${\pm}$23 & 226.7${\pm}$6.8 & 105${\pm}$12 & & & 2.03${\pm}$0.15  \\
$-$24 46 19.6 & & & & & & & &  \\
\\
$ $16 26 51.86 & & 25.84${\pm}$11.2 & 206${\pm}$8.5 & 52.3${\pm}$1.6 & & & & 2.43${\pm}$0.51  \\
$-$11 27 23.9 & & & & & & & & \\
\hline
\end{tabular}

{$^{\star}$} References: NED -- NASA Extragalactic Database; TGSS -- http://tgss.ncra.tifr.res.in ; WISH -- de. Breuck et al. (2002); NVSS -- Condon et al.\ (1998); PMN -- Griffith et al.\ (1994); ATCA -- Murphy et al.\ (2010)\\
$^{\#}$ For the detected sources the values given refer to peak flux densities, since these inverted spectrum sources are expected
to be compact, as also confirmed by their TGSS images.\\ 
For the undetected sources the quoted flux densities are 2.5$\sigma$ upper limits. All flux densities and errors have been recalibrated using the FSF values determined for the respective TGSS frames (Sect.2)\\ 
$^{\dag}$ In accordance with the practice followed by the authors of the WISH survey, De Breuck et al (2002), in their extensive spectral study, we have used integrated flux densities given in the WISH catalogue at 352 MHz.\\
{$^{\P}$}`Strong Cases' refer to the two EISERS for which $\alpha$ is more likely  to exceed +2.5, since the estimated $\alpha$ (150 - 352 MHz) is a lower limit which is close to the critical value $\alpha_c = $+2.5.\\ 

\end{table}
\end{landscape}

\clearpage

\newpage
\begin{figure}
\centering
\hspace*{-1.0cm}
\hbox{
\includegraphics[height=22.0cm,width=16.0cm]{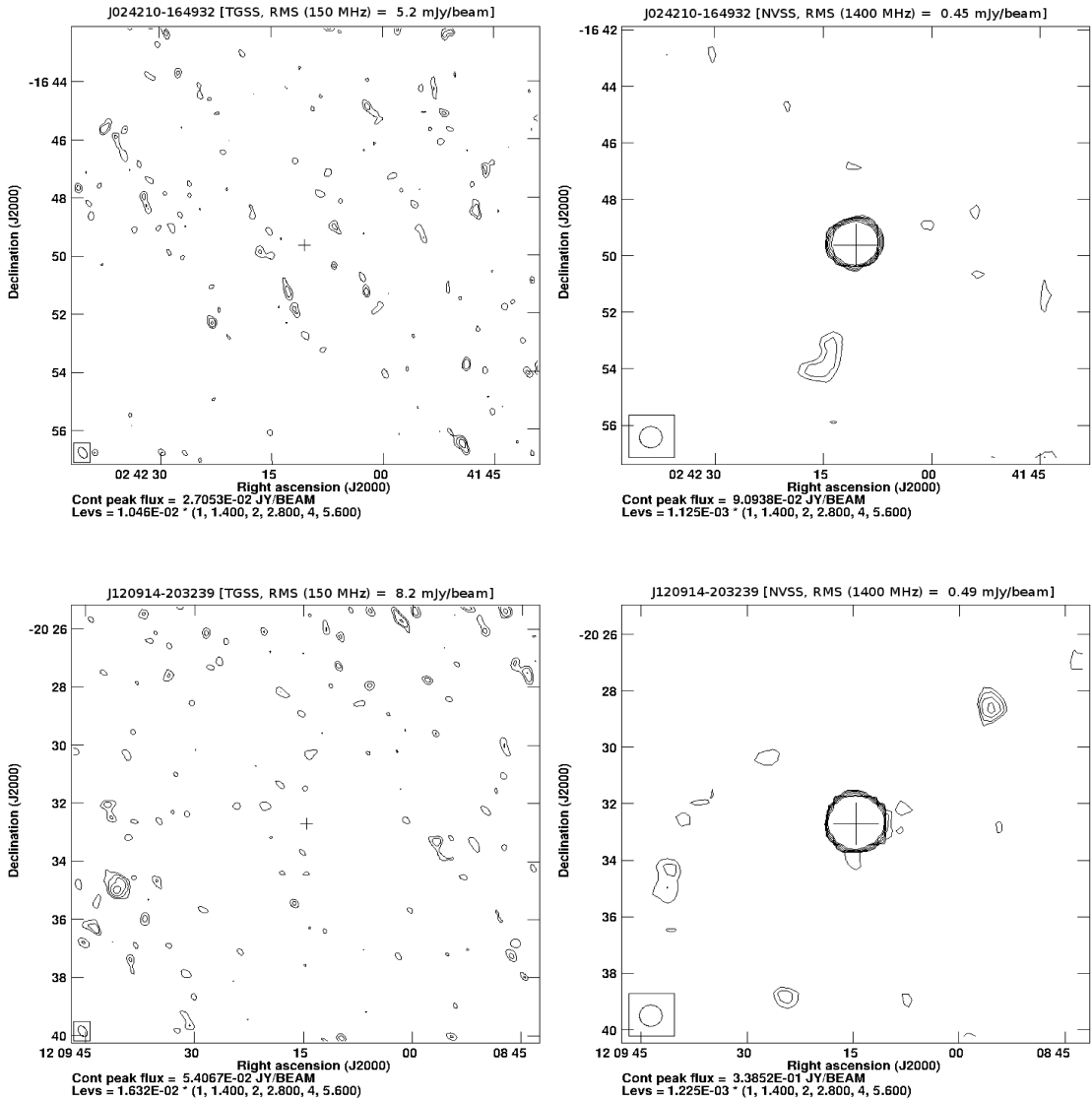}
}
\caption{Radio fields of the two EISERS undetected at 150 MHz (Table 1). Each row shows 15' x 15' wide radio maps at 150 MHz (TGSS) and 
1.4 GHz (NVSS), centered at the WISH position of the source marked with a `+' sign.}  
\label{fig:1}
\end{figure}
\clearpage

\newpage
\begin{figure}
\hspace*{-1.0cm}
\hbox{
\includegraphics[height=22.0cm,width=16.0cm]{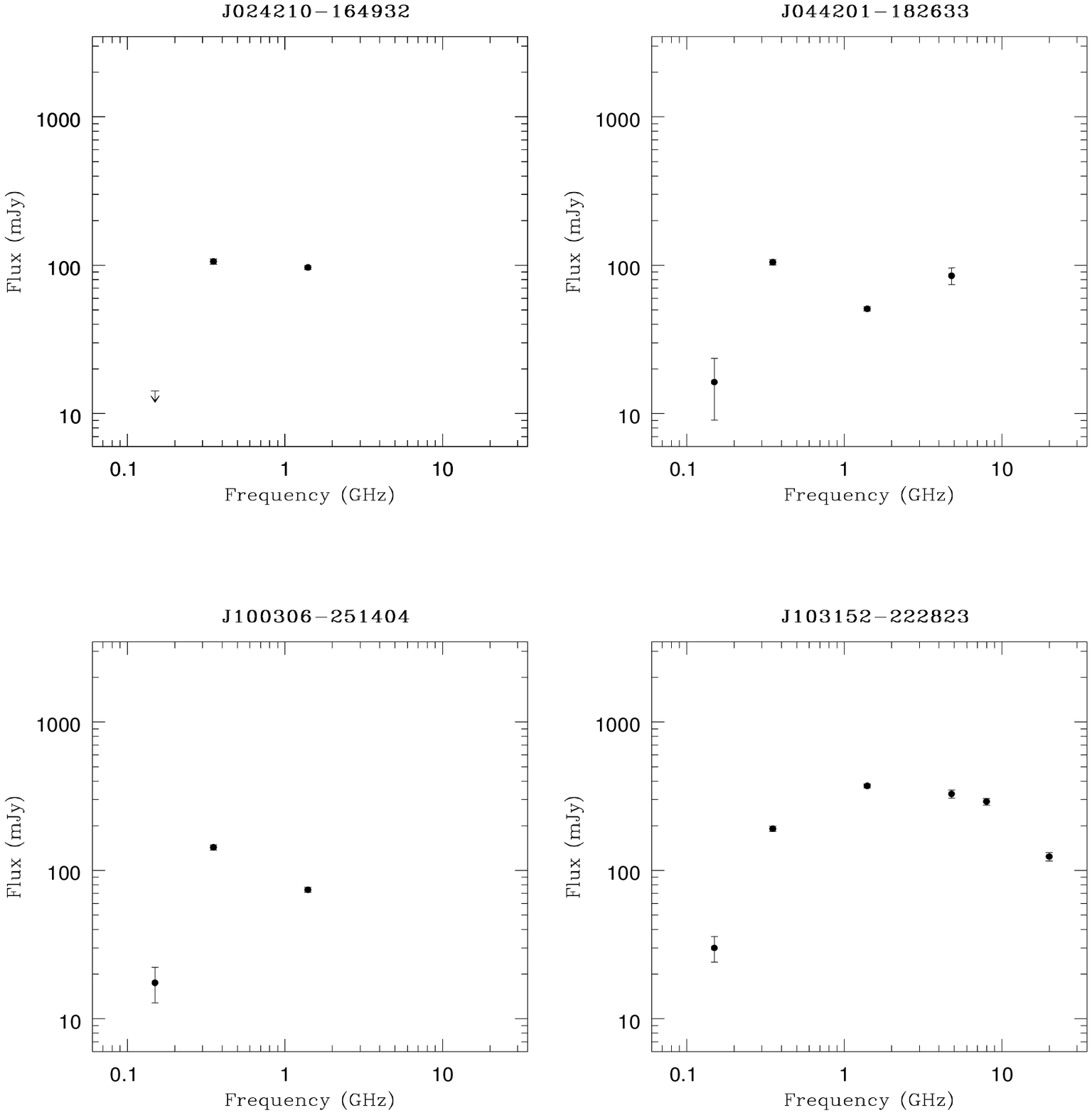}
}
\caption{Radio spectra of the seven EISERS (Table 1).}
\label{fig:2}
\end{figure}
\clearpage

\newpage
\begin{figure}
\hspace*{-1.0cm}
\hbox{
\includegraphics[height=22.0cm,width=16.0cm]{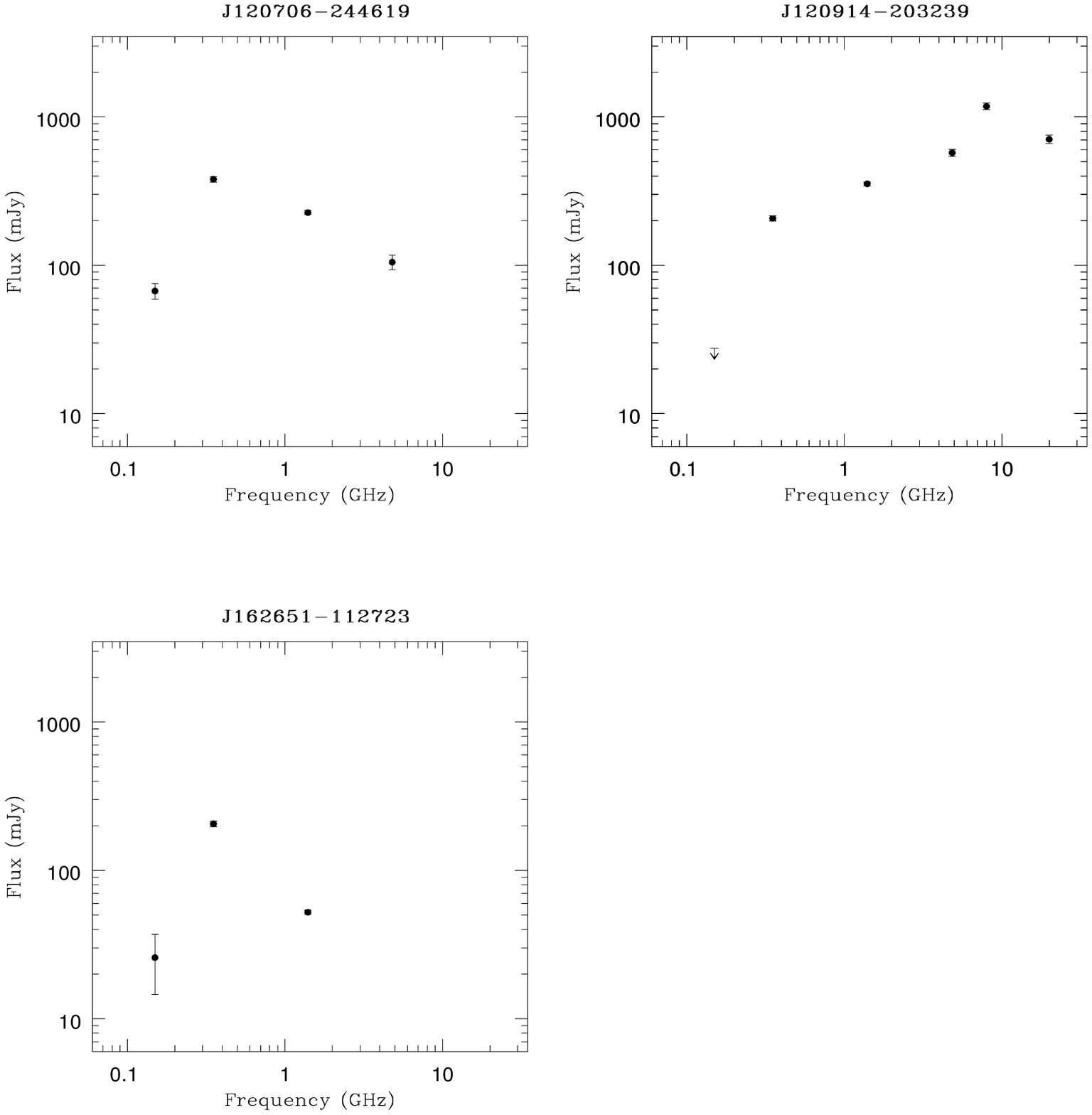}
}
{{\bf Figure~\ref{fig:2}}. \textit {continued}}
\end{figure}
\clearpage
\end{document}